# Two-electron quantum walks can probe entanglement and decoherence in an electron microscope

Offek Tziperman[1], David Nabben[2], Ron Ruimy[1], Jacob Holder[2], Ethan Nussinson[1], Yiqi Fang[2], Alexey Gorlach[1], Daniel Kazenwadel[2], Aviv Karnieli[3], Ido Kaminer[1*], Peter Baum[2*]

[1] *Solid State Institute, Technion – Israel Institute of Technology, 32000, Haifa, Israel*
[2] *Fachbereich Physik, Universität Konstanz, 78464, Konstanz, Germany*
[3] *E. L. Ginzton Laboratories, Stanford University, 94305, Stanford CA, USA*

[*] *Corresponding authors. Email: kaminer@technion.ac.il, peter.baum@uni-konstanz.de*

**Classical physics is often a good approximation for quantum systems composed of many interacting particles, although wavepacket dispersion and scattering processes continuously induce delocalization and entanglement. According to decoherence theory, an entangled ensemble can appear classical when only a subset of all particles is observed. This emergence of macroscopic phenomena from quantum interactions is, for example, relevant for phase transitions, quantum thermalization, hydrodynamics, spin liquids, or time crystals. However, entanglement and decoherence in free electrons have not yet been explored, although the electron is a fundamental elementary particle with extraordinary technological relevance. Here, we investigate the degree of coherence and entanglement in a free-space electron gas in the beam of an ultrafast electron microscope. We introduce a two-electron quantum walk that transforms the quantum state into different bases for quantum state tomography of entangled or partially entangled electron-electron pairs. We apply this novel diagnostic to study quantum effects in short pulses of hundreds of electrons under strong Coulomb correlation. We observe a high contrast interference in the electron-electron correlations but no significant signs of electron-electron entanglement which we explain by limited purity of the initial states and decoherence effects from unmeasured reservoir electrons. The ability to characterize quantum states of multiple free electrons may allow verification of electron-electron entanglement for use in fundamental studies and quantum electron microscopy.**

According to the laws of quantum mechanics, an ensemble of many interacting particles evolves coherently, and the corresponding unitary evolution delocalizes the wavefunction of each particle while also entangling it with all other particles in reach[1-3]. For example, two delocalized electrons interacting by Coulomb forces produce entanglement[4] because different parts of each wave function simultaneously interact with all parts of the other one. A multi-particle system with strong interactions, therefore, should develop a highly multidimensional and widespread entanglement. However, in practice, only a subset of all particles is observed, and the rest can be considered as an environment that takes away quantum information. In such a scenario, decoherence theory predicts the appearance of classical properties[1-3]. This transition between microscopic quantum interactions and emergent macroscopic phenomena is, for example, relevant for phase transitions[5-8], quantum thermalization[9], hydrodynamics[10,11], spin liquids[12], or time crystals[13].

A so far unexplored entity in this context is the free electron, one of the most ubiquitous and simple elementary particles in atoms, molecules, condensed matter, and materials. In contrast to photons, electrons carry a charge, and in contrast to atoms or ions, their mass is low. Consequently, multiple electrons interact strongly and directly with each other through Coulomb forces and provide an opportunity for studying multi-particle entanglement. Electrons also interact efficiently with coherent laser light, providing diagnostic tools[14-17]. For example, free electrons can interact with photons in a quantum-mechanical way[18-20]. Multiple free electrons can have classical anti-correlations in energy and time[21-23,45], photoelectrons from atoms can be entangled with the remaining ions[24,25], and electron pairs from laser-ionized $H_2$ molecules can produce coherent interference effects[26]. Recent groundbreaking works have proved the entanglement of free electrons and photons[27,28]. However, the quantum properties of multiple free-electrons and their degree of entanglement have remained unexplored, although theoretical studies suggest such possibilities[18,29,30].

**Question and Experimental Design**

The main idea of our theory and experiment is depicted in Figure 1. We consider electrons in free space with continuous variables in the time-energy domain. There are three limiting cases: Figure 1a shows two delocalized and entangled electrons with time-energy correlations that defy a classical description. Figure 1b depicts two delocalized electrons with extended wavefunctions, without entanglement but with anti-correlations in energy and time. Figure 1c shows two classically anti-correlated point-like electrons[21-23]. In the experiment (Fig. 1d), we investigate the delocalization and entanglement in such electrons with help of an ultrafast electron microscope[31]. We apply femtosecond laser pulses at a sharp metallic needle tip to create a dense electron gas with about one hundred electrons per pulse. These electrons then interact with each other via Coulomb forces (upper inset) that can create and destroy entanglement[4]. Later, an aperture selects one electron or a pair of electrons from the entire gas to be investigated for remaining delocalization and entanglement.

Measuring the quantum-mechanical properties of a two-electron quantum state requires a series of measurements in multiple bases of the two-electron subspace, accessible by suitable coherent transformations of the two-electron state in the continuous variables under scrutiny, here time and energy. For this purpose, we use a series of apertures to produce an almost linear beam of two-electron states with minimized spatial properties and then apply laser pulses (red) that modulate the energies of both electrons simultaneously in a coherent and controlled way. The laser wavelength is $\lambda = 1030$ nm, the pulse duration is 1.2 ps, the photon energy is $E_{ph} \approx 1.2$ eV and the pulses are long enough to cover all potential electron arrival times[23]. The resulting two-electron coincidence maps are then measured by energy-resolved detection on a number-counting screen.

Inspired by earlier results on the laser-modulation[14-17] and quantum state tomography of single electrons[32], we design our experiment to produce a two-electron quantum walk (Fig. 1d, lower inset). At an ultrathin silicon nitride membrane, the two-electron wavefunction is periodically accelerated or decelerated by the optical electric field, and the two-electron state becomes modulated in energy. The coupling parameter $g$ is proportional to the product of electric field strength and effective membrane interaction time[16]. Our laser pulses cover the entire two-electron wave function in space and time, and the photons therefore mediate coherent and correlated transitions of our two electrons between different energy states (Fig. 1d, lower inset). Final energies can be reached through multiple and interfering two-electron paths, providing insight into coherence and entanglement of the original two-electron state.

**Results**

Figure 2a shows the measured effective electron energy spectrum as a function of the modulating laser power for one-electron and two-electron events. The horizontal axis shows the energy shift $\Delta E$ compared to the zero-loss peak in units of laser photon energy $\hbar\omega$. At zero modulation power, the electron spectrum of the one-electron events is a single peak at $\Delta E = 0$ with a width of 1.15 eV (left column). Under the same conditions, the spectrum of the two-electron events has two peaks that are separated by about 1.5 eV (right column), caused by a classically explainable anti-correlation between the two electrons due to Coulomb repulsion near the emitter tip[21-23]. When we increase the laser power on the modulation membrane, we observe a broadening into multiple additional sideband peaks, accompanied by power-dependent oscillations in each peak's amplitude, indicating the Rabi revivals of a coherent quantum walk[14,17]. In the one-electron case (left column), we reproduce earlier results by Feist et al.[17] whereas the two-electron walk (right column) shows a different shape. In contrast to the one-electron data (left panel) in which the zero-loss peak is initially maximum and then decreases, the zero-loss peak in the two-electron case (right panel) has initially a minimum and only obtains intensity at larger laser fields. Figure 2b shows effective spectra at selected modulation strengths, highlighting the differences between one-electron and two-electron events. The bottom panels of Fig. 2a show theoretical results, obtained by using the measured energy spectra with no laser for one or two electrons per pulse followed by convolution with single-electron theory[14,15,20] (methods). We see a good match with the experiment.

Fringe visibility in energy-domain electron interferometry quantifies the ability of the de Broglie wave of an electron to interfere with itself in the time domain. Electrons with low temporal coherence do not cover multiple optical cycles of the modulating laser wave in a coherent way and cannot produce distinguishable energy sidebands via spectral interference effects[33]. In Fig. 2b, top panel, we read a fringe visibility of 70% for one-electron events but only 17% for two electrons. It thus might appear that our measured effective electron spectrum of the two-electron events originates from pairs of single electrons that are each localized in the time domain and have lost the necessary coherence length to produce coherent spectral interference effects in the quantum walk. However, we need to distinguish between single-particle coherence and multi-particle effects. Figure 2c shows measured coincidence maps $P(E_1, E_2)$ of our two-electron events. At 0 mW of modulation power (upper left panel), we observe classical anti-correlation in the energy domain from Coulomb effects[21,22]. The clarity of these anti-correlations indicates that the pairs of electrons that are measured have interacted strongly and directly with each other[21,22]. At higher laser powers, we observe the onset of an intricate pattern of constructive and destructive interferences in the form of multiple circular peaks. The high fringe visibility of 68% for 4 mW of laser power indicates that our two electrons largely maintain their combined coherent wave-like features despite their noticeable Coulomb interaction dynamics in the gas. Although the overall two-electron spectrum in Fig. 2b becomes blurred, the coincidence measurements in Fig. 2c show that the two electrons do not localize into point-like particles but maintain their ability to interfere coherently with multiple cycles of laser light.

**Entanglement**

What is the quantum state of the electron pairs emerging from the gas, and are they entangled? For the following analysis, we consider a two-electron density matrix $\rho$ that is obtained from the full density matrix of the electron gas by tracing out all its unobserved parts, that is, all unobserved electrons and also the unobserved properties of each electron pair, such as transverse spatial coordinates and spin. We obtain $\rho(E_1, E_2, E_1', E_2')$, where $E_1, E_1'$ and $E_2, E_2'$ are the energies of the two electrons as a continuous variable on no particular grid. Without laser modulation, the measured probabilities to find the electrons at energy combinations $E_1, E_2$ are provided by the diagonal elements $P(E_1, E_2) = \langle E_1, E_2|\rho|E_1, E_2\rangle$ of $\rho$. Access to the off-diagonal elements is obtained by coherently transforming the two-electron state with our laser interaction, followed by measuring $P(E_1, E_2, g)$ and fitting the data to theory with a maximum-likelihood density matrix reconstruction approach[34-36]. Since our scheme measures the entanglement of electrons generated prior to the laser modulation, rather than relying on particle indistinguishability, we do not need to overlap the electrons' full Hilbert space in the experiment.

Let us derive the laser-modulated coincidence spectrum for a general two-electron quantum state which may or may not be entangled. First, we represent the density matrix as an ensemble of pure states $\rho = \sum_n \lambda_n |\psi_n\rangle\langle\psi_n|$, where $\lambda_n$ are the probabilities associated with each state $|\psi_n\rangle$ in the two-electron ensemble. Subsequently, we derive the two-electron spectrum for a general pure state $|\psi\rangle$ (methods):

$$P(E_1, E_2, g, \psi) = \sum_{n_1, n_2, m_1} J_{n_1}(2|g|) J_{n_2}(2|g|) J_{m_1}(2|g|) J_{n_1+n_2-m_1}(2|g|)$$
$$\psi(E_1 - n_1\hbar\omega, E_2 - n_2\hbar\omega)\psi^*(E_1 - m_1\hbar\omega, E_2 - (n_1 + n_2 - m_1)\hbar\omega). \quad (1)$$

Here, $g$ is the laser-electron coupling strength[17] and $J_k$ denote Bessel functions of the first kind. The indices $n_1, n_2$ and $m_1$ are related to the number of laser photons that are emitted or absorbed by the first or second electron (see Fig. 1d, lower inset). The two-electron coincidence spectrum for an arbitrary mixed state is then $P(E_1, E_2, g, \rho) = \sum_n \lambda_n P(E_1, E_2, g, \psi_n)$.

We see that the measurable spectra $P(E_1, E_2)$ depend in a non-trivial way on the two-electron density matrix $\rho$ and the coupling strength, providing opportunity for reconstruction. However, the Hilbert space of our two free electrons is very large. If we would discretize the energy axis into 100 bins, corresponding to an energy resolution of approximately 100 meV, the density matrix would contain $100^4 = 10^8$ complex parameters. Therefore, we constrain our search for a $\rho$ that best matches the experiment to a subspace. We model the two-electron quantum state $\rho$ as a mixture of a fully entangled state $\psi_{\text{ent}} = \sqrt{P(E_1, E_2)} e^{i\phi(E_1, E_2)}$ and a separable state $\rho_{\text{sep}}$ with the same diagonal elements. The total state is expressed as $\rho(f, \phi) = f|\psi_{\text{ent}}\rangle\langle\psi_{\text{ent}}| + (1 - f)\rho_{\text{sep}}$ where $f$ is the entanglement fraction. We parametrize the phase $\phi$ of the two-electron wavefunction as $\phi(E_1, E_2) = (E_1 - E_2)(\alpha + \beta E_1/2 + \gamma E_2/2) + G(E_1 + E_2)$, where $\alpha, \beta, \gamma$ are linear and quadratic phases as the first orders of dispersion that are likely to emerge from Coulomb interactions between two extended electron waves. The phase $G(E_1 + E_2)$ is a general function of $E_1 + E_2$ that characterizes, for example, temporal separation of the electrons after moving with differing velocities[23]. Interestingly, $G$ does not affect any results (methods), and the two-electron quantum walk is therefore robust to this kind of temporal separation.

We can now fit the experimental results and find out which of the three regimes in Fig. 1 matches best with our experiment. Figure 3a shows a classical simulation of point-like particles (methods). In Figs. 3b-d, we plot theoretical two-electron coincidence spectra with entanglement ($f = 1$) for selected phases $\phi(\alpha, \beta, \gamma)$. Figure 3e shows the simulated two-electron coincidence map for no entanglement ($f = 0$). For comparison, Fig. 3f shows the measured results.

From the qualitative match to the $f = 0$ case, we conclude that the electrons are not much entangled, but rather correlated matter waves described by a separable state. A fit of our theory to the experimental data with a global fit of $f, \alpha, \beta, \gamma, g$ to all measured correlation maps at all laser powers delivers an optimum at $f = 0.18$ with phases of $\alpha = 0.91/(\hbar\omega)$ and $\beta \approx \gamma \approx 0/(\hbar\omega)^2$ at a negativity of entanglement of 7% (see methods), but the deviation from a separable state is not significant (see Extended data Fig. S1). Within the limits of our experimental sensitivity and how we search in the available Hilbert space, the two-electron quantum state is a non-entangled but coherent matter wave (Fig. 1b). We see that the reported two-electron quantum walk can distinguish between these cases and quantify the blend.

**Decoherence and loss of entanglement**

How is entanglement formed and lost in the electron gas? We measure the single-electron coherence length and compare it with the average distance between different electrons before selection of the measured pair (see Fig. 4a). Entanglement is expected when the size of the single-electron wavefunctions is comparable to their average separation, while a rather classical anti-correlation should emerge when the mean particle distance is much larger than the coherence length. Figure 4b shows a measurement of the single-particle coherence length using the photon-order-peak-shift method[33,37]. By fitting the experimental results (Fig. 4b) to theory (Fig. 4c), we obtain a temporal coherence of 4.7 fs (full width at half maximum). Space charge effects reduce this coherence, and the measurement is thus a lower limit on the temporal coherence directly at the tip.

The mean distance between the electrons is obtained by comparing measured energy spreads of single electrons due to Coulomb repulsion (the Boersch effect) with a classical ab initio point-particle simulation of our emitter geometry. Figure 4d shows the results. The match between the measured energy spread (diamonds) and the simulation data (dots) provides a calibration of the absolute number of electrons that are emitted from the tip. We have 135 electrons per pulse (dashed line). Figure 4e shows a snapshot of the electron gas taken 400 fs after emission from the tip. The gas is approximately as long as it is wide, so the Coulomb forces affect the time-energy domain. Consequently, electrons in front are pushed forward and have higher relative energy (red dots) than electrons behind (blue dots). Figure 4f shows the rate of spectral broadening (solid line) and the central energy (dotted line) as a function of propagation time. We observe that the electrons acquire almost all of their final spectral width within ~400 fs, when they are still slow and close. Figure 4f shows the development of the average electron-electron distance as a function of propagation time. In the time range that is most relevant for Coulomb interactions and potential creation of entanglement (0-400 fs), the median nearest-neighbors distance (dashed line) is ~300 nm. This corresponds to ~30 fs in the time domain, which is only about six times larger than the lower limit of the electron coherence time. These results indicate that the total multi-electron wave function before selection of the pairs should have a certain amount of entanglement.

However, the type and phase of entanglement may depend on fluctuations in the gas. Also, we only measure two out of many electrons, and the rest can be considered as an environment[1-3] that carries away quantum information from the two-electron state. The unobserved electrons in the gas cause decoherence effects and create an apparent loss of entanglement between the two electrons under scrutiny. Wavepacket coherence is less susceptible to these effects and therefore persists, in alignment with our observation of substantial interference in the two-electron correlation maps (Fig. 2) in contrast to a non-significant entanglement (Fig. 3). In view of these thoughts, we argue that the electron gas in our experiment is an example of a strongly interacting quantum system in which Coulomb collisions create entanglement that is later lost by decoherence phenomena, providing insight into a regime in which classical physics emerges from quantum-mechanical principles (see Fig. 4a).

**Outlook**

The reported results show that a two-electron quantum walk in coherent laser pulses can provide insight into the coherence properties and potential entanglement between multiple free-space electrons. An analysis of measured two-dimensional correlation shapes for multiple interaction strengths provides the necessary basis change for quantum state tomography and future verification of entanglement. Our method should allow to characterize any multi-electron state that can efficiently interact with laser light, including low-energy electrons[38], electrons in two-color laser fields[39] or photo-emitted electrons from the inside of correlated materials. The electrons do not need to be well-controlled quantum states but can originate from semi-statistical sources such as an electron gas. Our quantum tomography can be extended to density matrix reconstructions of three or more particles for directly seeing the reduction of entanglement by selection of reduced number states. A production of significantly entangled electron-electron pairs seems in principle possible by utilizing Coulomb interactions in a focus of a dense electron beam. According to the numerical results, the use of shorter and weaker photoemission pulses on a narrow-band emitter tip will maximize coherence length and gas density for strong Coulomb interactions while minimizing the overall number of electrons for low decoherence effects. Such entangled electron pairs would not only be useful for low-dose electron microscopy[40,41], ghost imaging[42,43], or quantum electron microscopy[40,41] but also for further fundamental studies of how classical phenomena emerge from quantum principles in the realm of massive particles with electric charge.

Looking at the big picture, we argue that free electrons are an ideal platform for studying many-body quantum physics and the creation and loss of entanglement. Several unique characteristics of electrons set them apart from other entities like trapped atoms or ions, photons, or superconducting circuitry. Coulomb forces provide direct long-range interactions, electron beams can be controlled on ultimately small dimensions in space and time, multi-electron states can be coherently controlled with laser light, and electrons can be measured with true number-resolving precision, features that are not simultaneously available in other schemes. We therefore argue that exploring many-body quantum physics with free electrons, now possible with the reported two-electron quantum walk, will provide new avenues for understanding the foundations and exploring the consequences of quantum mechanics in previously inaccessible regimes.

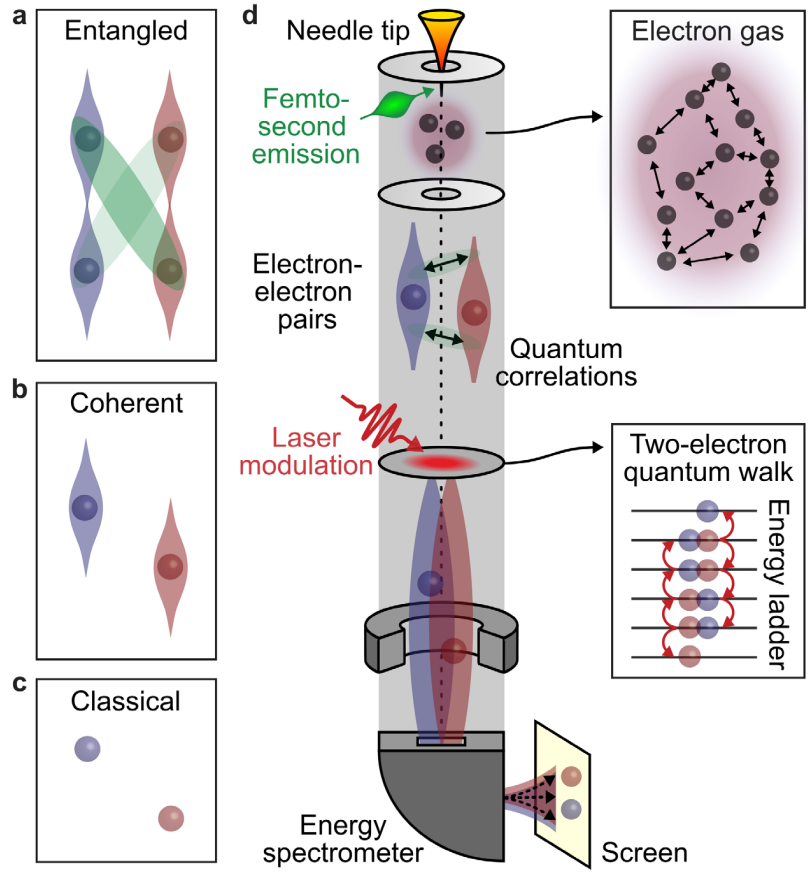

**Fig. 1. Concept and experiment for probing electron-electron entanglement and decoherence with a two-electron quantum walk.** We compare three limiting regimes of electron pair states: **(a)** Entangled electrons. **(b)** Classically correlated electrons with extended wavefunctions. **(c)** Classically correlated point particles. **(d)** Experimental setup. A femtosecond laser pulse (green) creates a dense gas of free electrons with strong Coulomb interactions (upper inset). We then select electron-electron pairs with unknown quantum correlations. For diagnostics of the two-electron quantum state, we let it interact coherently with a long modulation laser pulse (red). In a two-electron quantum walk (lower inset), pairs of electrons (blue, red) raise or lower their energy by integers of the photon energy. Finally, the resulting two-electron coincidence maps are measured with an energy spectrometer and an electron-counting screen.

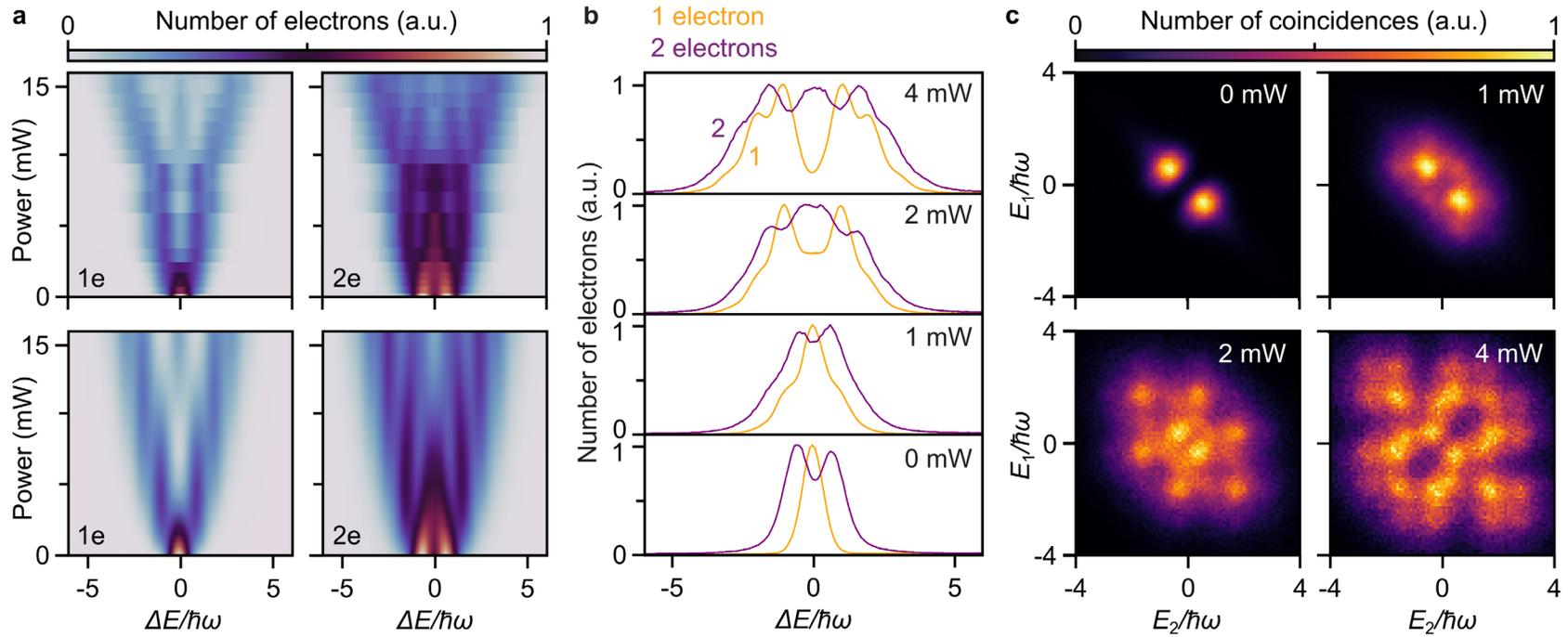

**Fig. 2. Two-electron quantum walk of electron pairs. (a)** Measured effective electron spectra as a function of laser power (upper panels) for one-electron (left panels) and two-electron events (right), each compared with theory (bottom panels). Intensity oscillations of the sideband peaks for one-electron and two-electron events are out of phase in terms of intensity. **(b)** Measured effective spectra for one-electron events (yellow) and two-electron events (violet) for selected laser powers. Fringe visibility appears higher for the one-electron events. **(c)** Measured two-electron coincidence maps after our two-electron quantum walk for selected laser powers. Each final pair of $E_1$ and $E_2$ can be reached in multiple ways, producing constructive and destructive interference that highlight the coherent nature of the two-electron quantum walk.

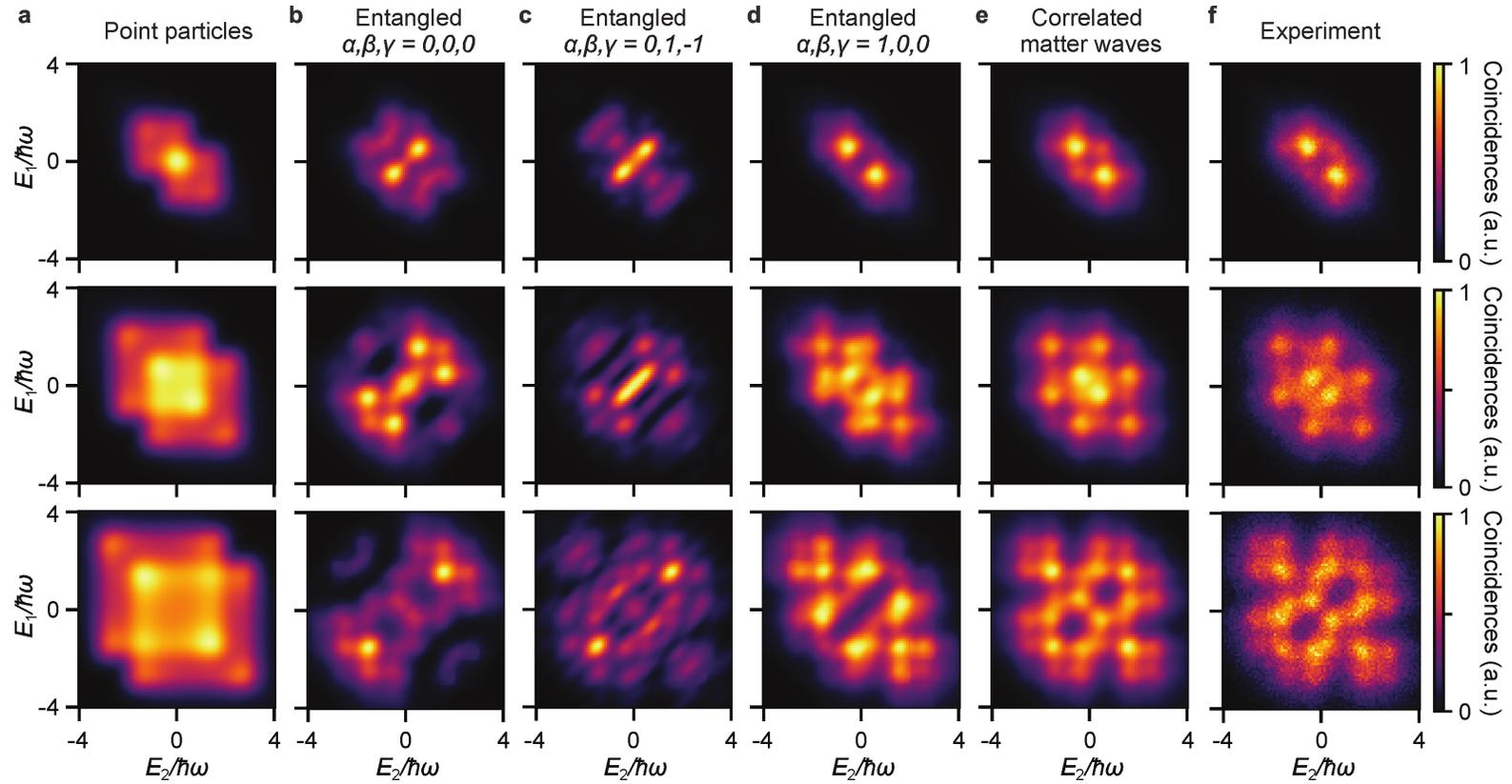

**Fig. 3. Two-electron quantum tomography in theory and experiment. (a)** Theoretical two-electron probability maps for point particles. **(b)-(e)** Theoretical two-electron probability maps for fully entangled electrons with different two-electron phases and separable matter waves with no entanglement. Top to bottom, 1 mW, 2 mW, and 4 mW, respectively. **(f)** Experimental results. We find that our measured electron pairs do not have a significant degree of entanglement and are best described by correlated matter waves.

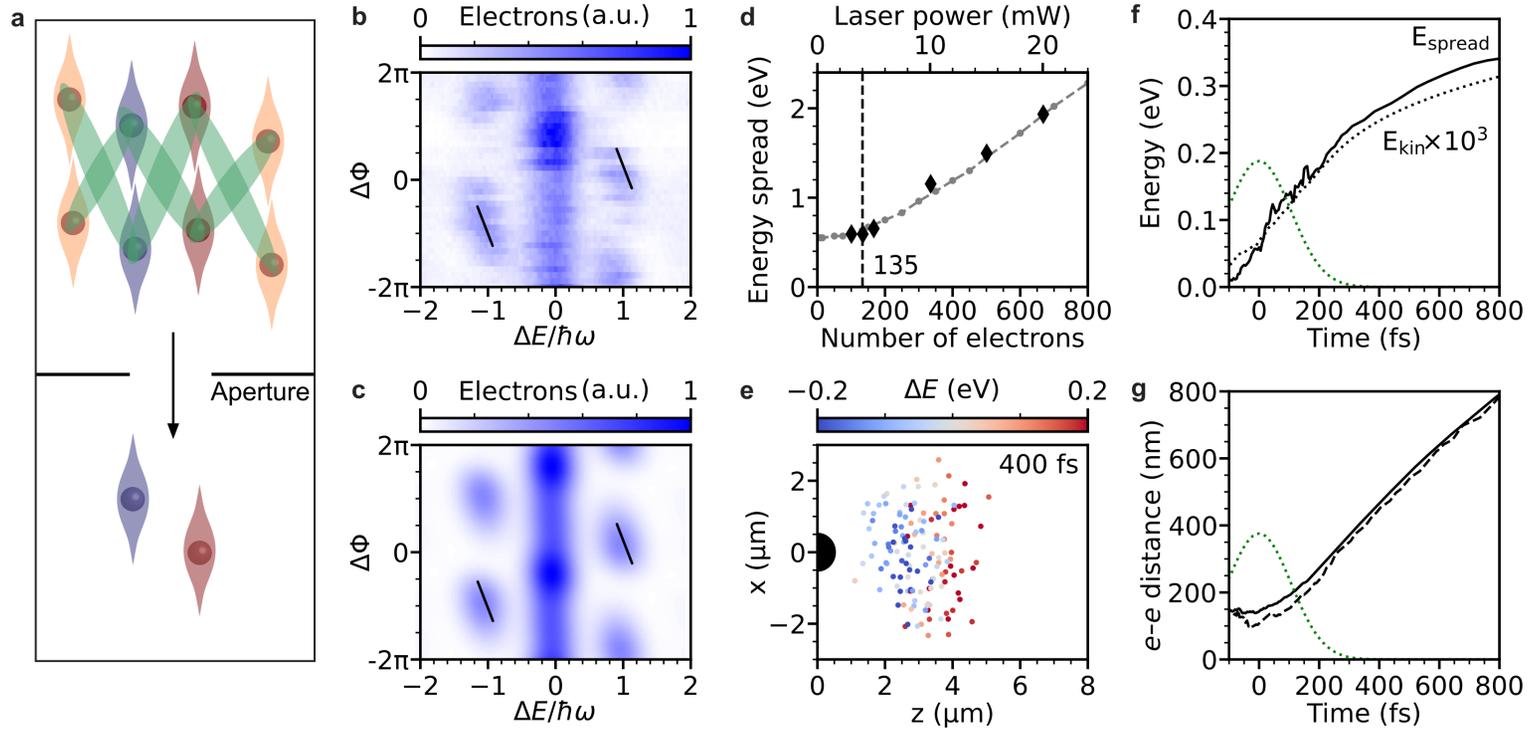

**Fig. 4. Loss of entanglement. (a)** Decoherence theory[1-3] predicts that an initially entangled multi-electron state of a dense electron gas (upper part) produces a subspace with a lesser degree of entanglement after the selection of only two electrons (lower part), because the unmeasured electrons act as an effective environment[1-3]. **(b)** A measurement of photon-order peak shifts (black lines) by attosecond electron microscopy[33,37]. **(c)** Fit results reveal for single electrons a temporal coherence length of 4.7 fs (full width at half maximum). **(d)** Comparison of measured energy bandwidth of single electrons (diamonds) and results from ab-initio point particle simulations (dots) of the emitter geometry reveal the number of electrons per pulse ($N$ = 135). Upper scale, power of the photoemission laser impinging at the needle tip. **(e)** Snapshot of the simulated electron gas at 400 fs after the emission time. Colors indicate the energy changes with respect to the electrostatic potential, caused by inter-electron Coulomb repulsion. The black half-circle depicts the needle tip. **(f)** Evolution of the spectral width (black solid line) and mean energy (black dotted line) of the simulated electron gas in relation to the emitting laser pulse (green dotted line). **(g)** Nearest-neighbor distance between two electrons as a function of propagation time. Solid line, average; dashed line, median; green dashed line, photoemission pulse. The electron-electron distance in the first 0-400 fs is 150-400 nm, comparable to the coherence length.

# Methods

**Experimental Details**: We use an ultrafast transmission electron microscope with a W/ZrO Schottky field emitter (JEM F200, JEOL), operated at an extraction voltage of $U_e$ = 1.65 kV and a filament current of $I_f$ =1.35A. A femtosecond laser (Carbide, LightConversion) at a wavelength of 1,030-nm provides 250fs long pulses at a 2-MHz repetition rate. A first beam is frequency-doubled with a beta-barium-borate crystal and then focused with an $f$ = 300-mm lens into a focus of ~20 μm diameter on the apex of the field emitter tip. The photoelectrons in the gas then interact by Coulomb forces while an electrostatic field accelerates them to a kinetic energy of 1.65 kV. After ~200 ps or ~2 mm, a small aperture absorbs most of the electron gas and isolates an electron beam with an average of only ~0.2 electrons per pulse, which we later categorize as zero-, one- or two-electron events. The electron beam is then further accelerated to a mean kinetic energy of 190 keV. The pulses in a second laser beam are stretched to 1.2-ps (full width at half maximum) by 7 passes through a 35-mm thick SF66 glass block. A gold-coated off-axis parabolic mirror with $f \approx$ 2.7 mm inside the electron microscope focused the beam to a diameter of ~8 μm at a 10-nm-thick silicon nitride membrane (Norcada) for modulation. Electron correlations after the two-electron quantum walk are measured with a post-column electron energy filter (CEFID, CEOS) and an event-based direct electron detector (Timepix3, Amsterdam Scientific Instruments) with real-time clustering[44].

**Time domain investigation of the two-electron phase space**: We characterize the two-electron dynamics in the time domain by sideband generation with less stretched laser pulses (~450 fs, full width at half maximum) on the modulation membrane. Extended data Fig. S2 shows the spectrum of single electrons or electron pairs as a function of time delay[45]. In analogy to Fig. 2, we observe in the single-electron spectrum a time-symmetric sideband pattern with destructive interference in the center. In contrast, the two-electron spectrum reveals a local maximum in the center and an asymmetric sideband spectrum with respect to time delay. The two electrons arrive at different times due to their different energies[23,45]. Extended data Fig. S3 shows the two-electron energy coincidence maps as a function of time delay. We observe a strongly asymmetric pattern where sidebands appear in varying directions at different delays. These results are consistent with measurements with even higher time resolution[44]. The fully stretched laser pulses (1.2 ps) cover the two-electron wave function completely (Fig. 3).

**Visibility:** Fringe visibility of the effective electron spectrum for single- and two-electron events (Fig. 2) is calculated as $V = (C_{\max} - C_{\min})/(C_{\max} + C_{\min})$, where $C_{\max}$ and $C_{\min}$ are the maximum and minimum electron count or coincidence values, respectively. Since $C_{\min}$ always approaches zero at large positive and negative energies, we restrict the energy range to $|E| < 2\hbar\omega$. For the coincidence maps of Fig. 2, we restrict both energy ranges $|E_1|, |E_2| < 2\hbar\omega$.

**Classical Coulomb Simulations**: To determine the number of electrons in the gas and their density, we perform quantitative classical simulations of electron propagation in the static fields of our emission geometry (Extended data Fig. S4a) with a voltage of 0 V at the tip, a suppressor voltage of -300 V, and acceleration voltages of 1.75 kV at A1 and 6.8 kV at A2. We obtain these static electric fields by a finite-element method (COMSOL Multiphysics). We initiate the photoelectrons at an initial kinetic energy of 1 eV normal to the surface and restrict the emission angle to 90°. The emission time is sampled from a Gaussian distribution with a width of 200 fs FWHM. The total number of electrons $N$ depends linearly on the laser power and is the only open parameter in our simulation. We numerically solve the equations of motion for all electrons in the electron gas in a classical molecular dynamics simulation. The scatter plot in Extended data Fig. S4a shows the simulated electrons' positions for different times. The calculated spectral widths agree with our measurements (Fig. 4d). Also, Extended data Fig. S4b-c compares simulated and measured spectral shapes for a large emitting laser power of 20 mW where space charge effects become nonlinear. Both spectra show significant Boersch effect[46,47] and two distinct peaks of matching shape and position. We thus conclude that the simulated numbers of electrons and gas densities (Fig. 4) match well to the experiment (Fig. 2-3).

**Theoretical fit for one-electron and two-electron quantum walks**: To fit theory to the laser modulated effective spectrum in Fig. 2a for selected one and two-electron events, we convolve the 0mW measured spectrum with Bessel amplitudes as in[15,20]:

$$P(\Delta E, g) = \sum_n J_n^2(2|g|) P(\Delta E - n\hbar\omega, g = 0). \tag{S1}$$

**Convolution for inhomogeneous $g$**: Our laser pulses are only ~2 times longer than the two-electron pulses in the experiment. In our entanglement analysis, we therefore average the theoretical maps over different values of $g$ according to the distribution of potential arrival times of the electrons within the optical field amplitude. We parametrize this distribution by using a nominal $g$ and the ratio of the electron's temporal width to the laser pulse width. The electrons position expectation values follow a normal distribution with standard deviation $\sigma_{el}$, and the laser amplitude is assumed to be Gaussian with a standard deviation $\sigma_{laser}$, calculated at the electron's position expectation value. We assume both electrons are transformed by the same laser amplitude. We find $\sigma_{el}/\sigma_{laser} = 0.48$, close to the independent measured ratio of pulse durations (see above). At these values, the distribution of $g$ is one-sided with a standard deviation of ~12%.

**Theory of two-electron coincidence for entangled or separable electron states:**

Pure states: We express a pure state, described by a two-electron wavefunction, as

$$\psi(E_1, E_2) = \sqrt{P(E_1, E_2, g = 0)} e^{i\phi(E_1, E_2)} \qquad (S2)$$

where the phase $\phi(E_1, E_2)$ is strictly real. For small deviations from the initial electron energy $\Delta E \ll E_0 \approx 190$ keV, the interaction of a single electron with the laser can be expressed by the scattering matrix $S(g) = e^{-i(g\hat{b}^\dagger + g^*\hat{b})}$, where $\hat{b}, \hat{b}^\dagger$ are energy ladder operators for a single electron, such that for the monoenergetic state of the electron $|E\rangle$ we have $\hat{b}|E\rangle = |E - \hbar\omega\rangle$ and $\hat{b}^\dagger|E\rangle = |E + \hbar\omega\rangle$; $g$ is the complex number that describes the laser-electron coupling strength[17]. The scattering matrix for two electrons is a tensor product of single-electron scattering matrices according to $S(g) = S_1(g) \otimes S_2(g)$. Given the initial state of electrons $\psi(E_1, E_2)$, the state after the interaction is

$$\psi_{\text{final}}(E_1, E_2) = \int dE_a dE_b\, \psi(E_a, E_b) |E_a, E_b\rangle \langle E_a, E_b | S_1(g) \otimes S_2(g) | E_1, E_2 \rangle. \qquad (S3)$$

Taking into account the orthogonality of the monoenergetic states, i.e., $\langle E_a | E_b \rangle = \delta(E_a - E_b)$, we obtain the wavefunction after the interaction:

$$\psi_{\text{final}}(E_1, E_2) = \sum_{n_1, n_2} J_{n_1}(2|g|) J_{n_2}(2|g|) e^{-i(n_1+n_2)\angle g} \psi(E_1 - n_1\hbar\omega, E_2 - n_2\hbar\omega), \qquad (S4)$$

where $\angle g = \arg(g)$ is the phase of $g$. To determine the coincidence spectrum, we take the absolute square of $\psi_{\text{final}}$ according to $P(E_1, E_2, g, \psi) = |\psi_{\text{final}}(E_1, E_2)|^2$ and obtain

$$P(E_1, E_2, g, \psi) = \sum_{n_1, n_2, m_1, m_2} J_{n_1}(2|g|) J_{n_2}(2|g|) J_{m_1}(2|g|) J_{m_2}(2|g|) e^{-i(n_1+n_2-m_1-m_2)\angle g} \times$$

$$\times \psi(E_1 - n_1\hbar\omega, E_2 - n_2\hbar\omega) \psi^*(E_1 - m_1\hbar\omega, E_2 - m_2\hbar\omega). \qquad (S5)$$

For fits to the experimental results, we average over different absolute phases of $g$, which is not locked between the photoemission and the quantum walk in the experiment, resulting in

$$P(E_1, E_2, g, \psi) = \sum_{n_1, n_2, m_1} J_{n_1}(2|g|) J_{n_2}(2|g|) J_{m_1}(2|g|) J_{n_1+n_2-m_1}(2|g|)$$

$$\psi(E_1 - n_1\hbar\omega, E_2 - n_2\hbar\omega) \psi^*(E_1 - m_1\hbar\omega, E_2 - (n_1+n_2-m_1)\hbar\omega). \qquad (S6)$$

General Mixed States: If the electrons are in a general mixed state, which may or may not be entangled, we can decompose their state into an incoherent mixture of pure states:

$$\rho = \sum_n \lambda_n |\psi_n\rangle\langle\psi_n|. \qquad (S7)$$

From the probability $P(E_1, E_2, g, \psi_n)$ for each pure state $\psi_n$ using Eq. (S6), we can extend the derivation above to a general mixed state:

$$P(E_1, E_2, g, \rho) = \sum_n \lambda_n P(E_1, E_2, g, \psi_n). \tag{S8}$$

Separable Quantum Electrons: For separable electrons, which are a specific case of Eq. (S7), the equations simplify. A general separable state is expressed as:

$$\rho = \sum_n P_n |\psi_n^A\rangle\langle\psi_n^A| \otimes |\psi_n^B\rangle\langle\psi_n^B| \tag{S9}$$

In this case, the laser-modulated coincidence spectrum is given by

$$P(E_1, E_2, g) = \langle E_1, E_2 | S(g)\rho S(g)^\dagger | E_1, E_2 \rangle = \sum_n P_n |\langle E_1, E_2 | S(g) | \psi_n^A, \psi_n^B \rangle|^2. \tag{S10}$$

Since $S(g) = S^A(g) \otimes S^B(g)$ we have:

$$\sum_n P_n |\langle E_1 | S^A(g) | \psi_n^A \rangle|^2 |\langle E_2 | S^B(g) | \psi_n^B \rangle|^2. \tag{S11}$$

We identify the terms $|\langle E_2 | S^B(g) | \psi_n^B \rangle|^2$ as the single-electron transition probability. Next, we average over the absolute phase of the laser field:

$$|\langle E_2 | S^B(g) | \psi_n^B \rangle|^2 = \sum_{k,l} J_k(2|g|) J_l^*(2|g|) \psi_n^B(E_2 - k\hbar\omega) \psi_n^{B*}(E_2 - l\hbar\omega) \cdot \langle e^{-i(k-l)\angle g} \rangle =$$

$$\sum_k J_k^2(2|g|) |\psi_n^B(E_2 - k\hbar\omega)|^2. \tag{S12}$$

The final probability is:

$$P(E_1, E_2, g, \rho) = \sum_{m,l} J_m^2(2|g|) J_l^2(2|g|) \sum_n P_n |\psi_n^A(E_1 - m\hbar\omega)|^2 |\psi_n^B(E_2 - l\hbar\omega)|^2. \tag{S13}$$

We can write this as a function of the coincidence map without laser modulation:

$$P(E_1, E_2, g) = \sum_{m,l} J_m^2(2|g|) J_l^2(2|g|) P(E_1 - m\hbar\omega, E_2 - l\hbar\omega, g = 0). \tag{S14}$$

We see that all separable states with the same initial probability map before interaction with the laser, $P(E_1, E_2, g = 0) = \langle E_1, E_2 | \rho | E_1, E_2 \rangle$, will have the same laser-modulated probability maps.

Classical electrons: In analogy to the supplement of Ref.[48], we assume an initial correlated energy spectrum of two point-electrons $P(E_1, E_2, g = 0)$. The electrons accelerate or decelerate by interaction with the laser's electromagnetic field depending on their arrival time. For a single monoenergetic electron, the classical spectrum for an energy change $\Delta E$ is:

$$P(\Delta E, g) \propto \frac{1}{\sqrt{4|g|^2(\hbar\omega)^2 - \Delta E^2}}, \tag{S15}$$

where $\Delta E$ is the final energy change from $E_0$, $\omega$ is the photon frequency, and $g$ is the laser-electron modulation strength[17]. The electron source in our experiments has a spectrum $P^{\text{el}}(E, g = 0)$ with finite bandwidth. The spectrum after interaction with the laser is given by the convolution of the initial spectrum with the spectral response:

$$P^{\text{el}}(E, g) \propto P(\Delta E, g) * P^{\text{el}}(E, g = 0). \tag{S16}$$

Substituting Eq. (S15) in Eq. (S16) gives:

$$P^{\text{el}}(E, g) \propto \int dE_0 \frac{1}{\sqrt{4|g|^2(\hbar\omega)^2 - E_0^2}} P^{\text{el}}(E - E_0, g = 0). \tag{S17}$$

The straightforward generalization for two electrons yields:

$$P^{\text{el}}(E_1, E_2, g) \propto \int dE_a dE_b \frac{P^{\text{el}}(E_1 - E_a, E_2 - E_b, g = 0)}{\sqrt{(4|g|^2(\hbar\omega)^2 - E_a^2)(4|g|^2(\hbar\omega)^2 - E_b^2)}}, \tag{S18}$$

which is the formula we use for Fig. 3; normalization factors are found numerically.

**Entanglement analysis by a maximum likelihood approach:** Here we explain the global fit of the experimental results based on measurements of the energy coincidence spectra with and without the two-electron quantum walk. We start with the two-electron density matrix $\rho = \sum_{E_1, E_2, E_1', E_2'} \rho(E_1, E_2, E_1', E_2')|E_1, E_2\rangle\langle E_1', E_2'|$. Our goal will be to estimate $\rho$, by utilizing the maximum likelihood approach with the available data. In the experiment without the laser modulation, we directly measure the diagonal elements of the two-electron density $P(E_1, E_2, g = 0, \rho) = \langle E_1, E_2|\rho|E_1, E_2\rangle$. The two-electron quantum walk coherently transforms the two-electron state into $P(E_1, E_2, g, \rho) = \langle E_1, E_2|S\rho S^\dagger|E_1, E_2\rangle$, where $P(E_1, E_2, g, \rho)$ is taken from Eq. 1 in the main text.

The subset of entangled states that we consider is an incoherent mixture of the state $\psi(\phi) = \sqrt{P_{\text{exp}}(E_1, E_2)}e^{i\phi(E_1, E_2)}$, $\rho_{\text{ent}} = |\psi(\phi)\rangle\langle\psi(\phi)|$ with a separable state $\rho_{\text{sep}}$ with the same diagonal elements corresponding to the measurement $P(E_1, E_2, g = 0) = \langle E_1, E_2|\rho_{\text{sep}}|E_1, E_2\rangle$. We parametrize the phase $\phi(E_1, E_2) = (E_1 - E_2)(\alpha + \beta E_1/2 + \gamma E_2/2) + G(E_1 + E_2)$, where $\alpha, \beta, \gamma$ are fit parameters and $G(E_1 + E_2)$ is a general function of $E_1 + E_2$.

The laser-transformed two-electron spectrum $P(E_1, E_2)$ is unaffected by any phase $G(E_1 + E_2)$. We prove this by calculating the spectrum for $\psi(E_1, E_2) = \sqrt{P_{\text{exp}}(E_1, E_2)}e^{i\phi(E_1, E_2) + iG(E_1 + E_2)}$ using equation S5 we find:

$$P(E_1, E_2, g|\psi) = \sum_{n_1, n_2, m_1} J_{n_1}(2|g|)J_{n_2}(2|g|)J_{m_1}(2|g|)J_{n_1 + n_2 - m_1}(2|g|)$$

$$\psi(E_1 - n_1\hbar\omega, E_2 - n_2\hbar\omega)\psi^*(E_1 - m_1\hbar\omega, E_2 - (n_1 + n_2 - m_1)\hbar\omega)$$
$$\exp(iG(E_1 - n_1\hbar\omega + E_2 - n_2\hbar\omega))\exp(-iG(E_1 - m_1\hbar\omega + E_2 - (n_1 + n_2 - m_1)\hbar\omega)). \quad (S19)$$

The phase factors in the last row cancel, and the phase $G(E_1 + E_2)$ does not affect the result. The experiment is therefore robust to correlated overall energies or times that could originate, for example, from drifts of the acceleration voltage or laser overlap in time.

The overall density matrix is then:

$$\rho(\phi, f) = f|\psi(\phi)\rangle\langle\psi(\phi)| + (1-f)\rho_{\text{sep}}. \quad (S20)$$

We calculate the laser-modulated spectrum for this family of density matrices and compute the log-likelihood. For a given density matrix $\rho$ and the measurement results $P(E_1, E_2, g)$, the log-likelihood is given by:

$$L_{\log}(\rho) \propto -\sum_{E_1, E_2}\left(Tr(S(g)^\dagger|E_1, E_2\rangle\langle E_1, E_2|S(g)\rho) - P(E_1, E_2, g)\right)^2. \quad (S21)$$

By assuming different density matrices, $\rho$, using equation 1 to find $P(E_1, E_2, g, \rho)$ and comparing to the experiment using Eq. $S21$, we find the likelihood of $\rho$. Then we vary the parameters that define $\rho$ to find the maximum likelihood density matrix. We use all measured laser-transformed states and fit all of them simultaneously.

To extract error values, we first fit the optimal parameters $\{\theta_i\}_{i=1}^N = \alpha, \beta, \gamma, f, \ldots$ Then, we calculate the gradient, and Hessian of the least squares loss function numerically via $(\nabla r)_i = \frac{\partial r}{\partial \theta_i}$, $h_{ij} = \frac{\partial^2 r}{\partial \theta_i \partial \theta_j}$, with the loss function $r = \sum_{E_1, E_2, g}\left(P_{\exp}(E_1, E_2, g) - P_{\text{theory}}(E_1, E_2, g)\right)^2$. We estimate the standard error in $r$ as $s^2 = \frac{r}{N-p}$, where $N$ is the number of data points and $p = 9$ is the number of parameters (see Eq S22). Finally, the covariance matrix $C$ is calculated as $C = h^{-1}s^2$. The standard errors are the square roots of the diagonal elements. The results are (errors are two standard deviations):

$$g(1\text{ mW}) = 0.56 \pm 0.003, g(2\text{ mW}) = 0.98 \pm 0.004, g(4\text{ mW}) = 1.33 \pm 0.006,$$
$$\sigma = 0.16 \pm 0.006 \text{ eV}, f = 0.18 \pm 0.01, \alpha = 0.91 \pm 0.005, \beta = 0.01 \pm 0.005,$$
$$\gamma = -0.04 \pm 0.01, \sigma_{\text{el}}/\sigma_{\text{laser}} = 0.48 \pm 0.01. \quad (S22)$$

Extended data Fig. S2 shows the residuals of these fits; panels a-c show data for 1 mW, 2 mW and 4 mW of laser power, respectively. On the left side, we compare the measured data to the fit results when considering potential entanglement. On the right side, we compare the measured data to the fit results when considering no entanglement. We observe an improvement for the entangled case compared to the separable case, as seen by the slightly smaller residuals, but the result is not significant. Remaining systematic shapes may be caused by deviations between the actual quantum state and the one in our ansatz since we did not search the full space of entangled states and phases. In addition, there may have been drifts of the reference spectrum $P_{\exp}(E_1, E_2, g = 0)$ during the measurement, wrongly detected three-electron events, or space charge effects between quantum walk and spectrometer. Also, the two electrons may have been transformed by slightly different laser amplitudes. In Extended data Figs. S2g-i, we subtract from the measured results 0.75 times the separable theory, which leaves a fraction with potentially high degree of entanglement that we can visually inspect for general shape. We observe qualitative similarities with fully entangled theory for the optimal parameters in equation S22 and $f = 1$, especially around the $E_1 - E_2 = 0$ axis, but a visual agreement does not imply significance.

**Negativity of entanglement:** Negativity of a quantum state is defined as[49,50]:

$$N(\rho) = \frac{||\rho^{\Gamma_A}|| - 1}{2}, \tag{S23}$$

where $\rho^{\Gamma_A}$ is the partial transpose of $\rho$ with respect to system $A$, and $||\cdot||$ is the trace norm. Equivalently, the negativity is the absolute value of the sum of the negative eigenvalues of $\rho^{\Gamma_A}$. For all separable states, all eigenvalues of the partial transpose are positive[48,49], and the negativity is zero. If the negativity is greater than zero, the state is therefore entangled. Furthermore, negativity is an entanglement monotone, meaning that local operations can only lower the negativity[48,49]. We calculate the negativity of the state $\rho = f|\psi_{\text{ent}}\rangle\langle\psi_{\text{ent}}| + (1-f)\rho_{\text{sep}}$ in our theory and experiment with $\psi_{\text{ent}}(E_1, E_2) = \sqrt{P(E_1, E_2, g = 0)}\, e^{i\phi(E_1, E_2)}$ by using its Schmidt decomposition:

$$\psi_{\text{ent}}(E_1, E_2) = \sum_n \sqrt{\lambda_n}\, \psi_n^A(E_1)\psi_n^B(E_2). \tag{S24}$$

In the same basis we can write $\rho_{\text{sep}} = \sum_n \lambda_n |\psi_n^A, \psi_n^B\rangle\langle\psi_n^A, \psi_n^B| \equiv \sum_n \lambda_n |n,n\rangle\langle n,n|$, where we define $|n,n\rangle \equiv |\psi_n^A, \psi_n^B\rangle$ to simplify notation and give us a simple basis in which to calculate the negativity. In the experimental data with the sample laser off, we find $\lambda_n = 0.751, 0.199, 0.020, 0.017, 0.006, \dots$. The density matrix and its partial transpose are then:

$$\rho = f \sum_{m,n} \sqrt{\lambda_m \lambda_n}\, |n,n\rangle\langle m,m| + (1-f)\sum_n \lambda_n |n,n\rangle\langle n,n|, \tag{S25}$$

$$\rho^{\Gamma_A} = f \sum_{m,n} \sqrt{\lambda_m \lambda_n}\, |m,n\rangle\langle n,m| + (1-f)\sum_n \lambda_n |n,n\rangle\langle n,n|. \tag{S26}$$

We truncate the density matrix at the third eigenvalue because the rest are small and dominated by experimental noise. The negativity is then:

$$N(\rho) = f \sum_{m>n} \sqrt{\lambda_m \lambda_n} \geq f\sqrt{\lambda_1 \lambda_2} \approx 0.072. \tag{S27}$$

For reference, the negativity for a 2-qubit Bell state is $1/2$.

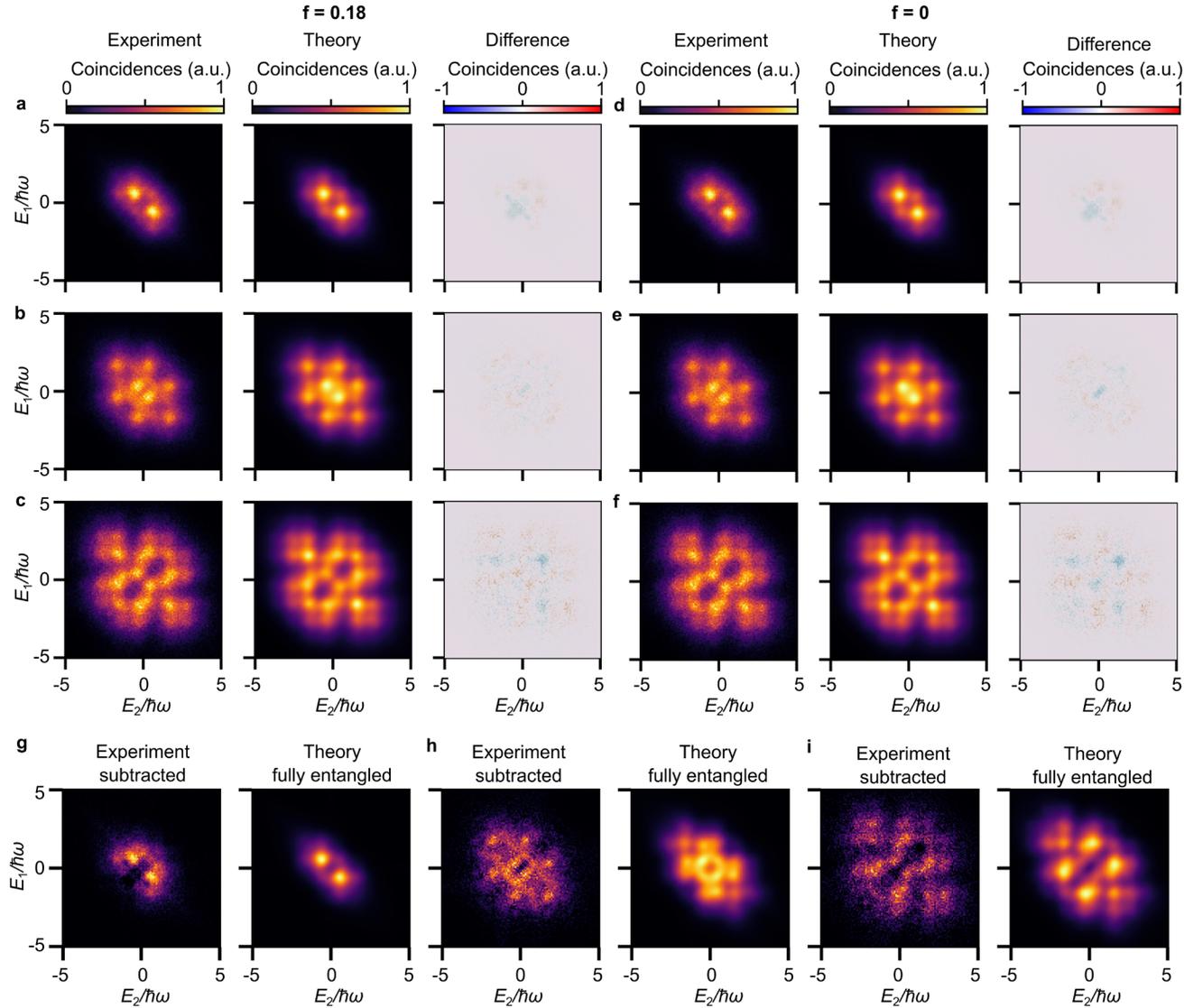

**Extended Data Fig. S1. Fit residuals. (a)-(c)** Measured data (left column) compared to best-fit theory considering entanglement (middle column) for laser powers of 1 mW, 2 mW and 4mW, respectively. The residual is small (right column). **(d)-(f)** Measured data (left column) compared to best-fit theory considering zero entanglement for laser powers of 1 mW, 2 mW and 4mW, respectively. The residuals (right column) are slightly larger and have a systematic shape. **(g)-(i)** Experimental data when subtracting 75% of separable (no-entanglement) theory. The residuals (left panels) compare slightly better to fully entangled theory (right panels) than to non-entangled theory (above), but overall, there are no significant signs of entanglement, and the state is best described by a coherent matter wave.

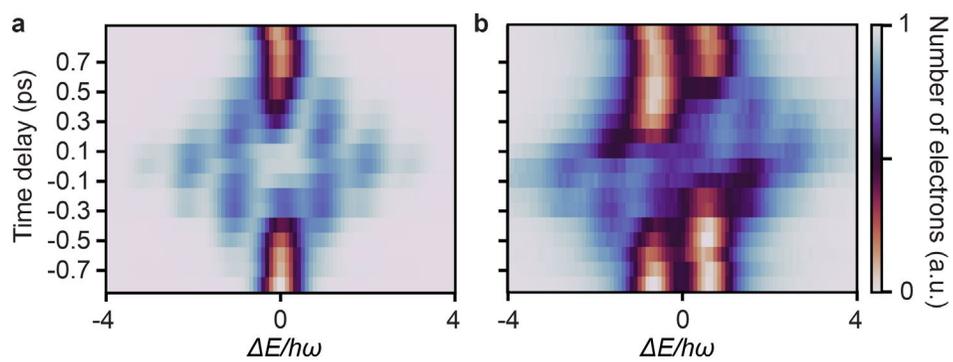

**Extended Data Fig. S2. Two electrons in the time domain.** (a) Time-resolved energy spectrum of single electrons in a non-stretched laser beam[45]. (b) Time-resolved energy spectrum of two-electron events.

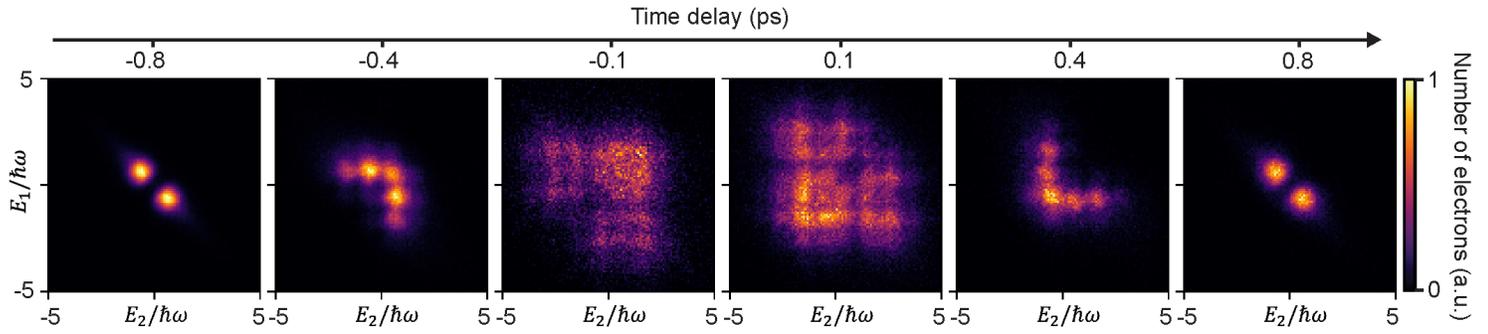

**Extended Data Fig. S3. Two-electron coincidence spectra resolved in time.** Measured electron coincidence maps as a function of time delay for shorter laser pulses (450 fs) that cover only part of the two-electron state. For large negative delays (second panel), mostly only the faster of the two electrons disperses into peaks. For large positive delays (fifth panel), mostly only the slower of the two electrons disperses into peaks. Compare Ref.[45].

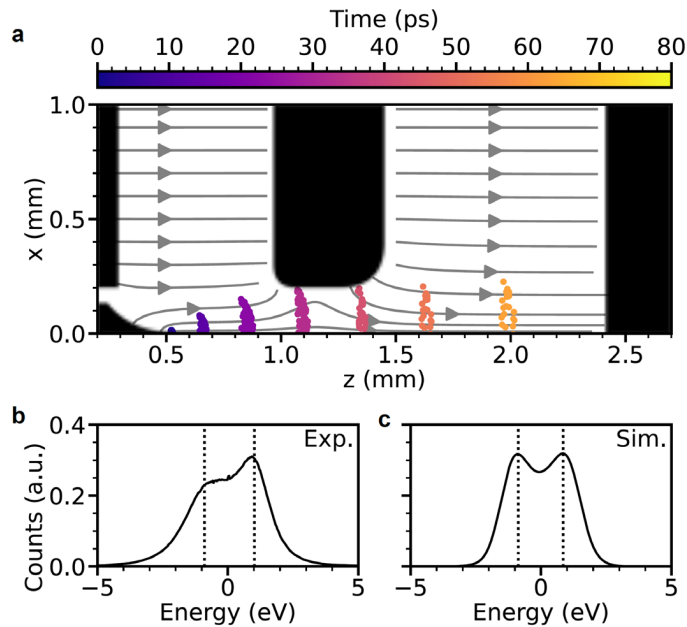

**Extended Data Fig. S4: Density of the electron gas. (a)** Geometry of the emitter section of our electron microscope (black) with calculated electric fields (grey). The scatter plot shows the electron distribution at different times after laser excitation (see color bar). **(b-c)** Measured and simulated electron energy distributions at an excitation power of ~20 mW. Both spectra show a similar amount of Boersch effect. The dotted lines show the interquartile widths.

**Acknowledgments:** We thank Rudolf Haindl for raw data on electric fields at needle tips[28] and Claus Ropers for helpful, critical remarks. This research was supported by the German Research Foundation (DFG) through SFB-1432. During preparation of this manuscript, we became aware of independent research on related topics[45].

**Author contributions:** OT and IK proposed the idea for the experiment. OT, RR, AG, EN and AK derived the theory. OT, DN and YF performed experiments and analyzed the data. JH and DK made the numerical point particle simulations. PB and OT wrote the manuscript with input from all authors.

**Competing interests:** The authors declare no competing interests.

**Data availability:** The data supporting the findings of this study are available from the corresponding author upon reasonable request.